\documentclass[aps,prd,twocolumn ,preprintnumbers,
superscriptaddress,nofootinbib]{revtex4}
\usepackage{graphicx}
\usepackage{amsmath,amssymb}
\usepackage{epsfig}
\usepackage{latexsym}
\newcommand {\ga} {\ {\raise-.5ex\hbox{$\buildrel>\over\sim$}}\ }
\newcommand {\la} {\ {\raise-.5ex\hbox{$\buildrel<\over\sim$}}\ }

\newcommand{\ba}{\begin{eqnarray}} \newcommand{\ea}{\end{eqnarray}}
\newcommand{\be}{\begin{equation}} \newcommand{\ee}{\end{equation}}

\usepackage{amsmath,amssymb,amsthm,graphicx,bm,color}
\usepackage{amsmath,amssymb,amsthm,graphicx,bm,color}

\renewcommand{\Delta}{\varDelta} 
\renewcommand{\Gamma}{\varGamma} 
\renewcommand{\Omega}{\varOmega} 
\renewcommand{\Phi}{\varPhi} 
\renewcommand{\Psi}{\varPsi} 
\renewcommand{\Sigma}{\varSigma} 
\renewcommand{\Theta}{\varTheta} 
\renewcommand{\epsilon}{\varepsilon}

\begin{document}

\title{Towards Realizing Warm Inflation in String Theory}
 
\author{Mar Bastero-Gil}
\affiliation{Departamento de Fisica Teorica y del Cosmos, Universidad de Granada, Granada-18071, Spain}

\author{Arjun Berera}
\affiliation{School of Physics and Astronomy, University of Edinburgh, Edinburgh EH9 3JZ, UK} 

\author{James B. Dent}
 
\author{Thomas W. Kephart}
\affiliation{Department of Physics and Astronomy, Vanderbilt
University, Nashville, TN 37235, USA} 

\date{\today}

\begin{abstract}
We give a generic argument that string theory provides a natural setting for warm inflationary cosmology. 
We then explore a specific model with an inflaton modulus field coupled to fields that provide the continuous dissipation needed for warm inflation and argue the results are generic for a large class of models.
\end{abstract}

\medskip


\maketitle

\vspace{-.5 cm}

The inflationary scenario has been very successful in explaining
various cosmological questions.  In the current era of precision
cosmological data, the physics governing inflation is becoming
increasingly illuminated \cite{Peiris:2003ff}, a situation which
should continue to improve with near future experiments
\cite{PLANCK:2006uk}.  On the theoretical side one would like to
demonstrate that inflation arises naturally from a fundamental theory.
The hope is that string theory can provide just such a microphysical
basis for inflation.   

There has been a substantial amount of recent work on embedding
inflation into string theory (for reviews and references see
\cite{Cline:2006hu}).  Much of this work is done within the landscape
of flux compactifications, focusing heavily on models in warped
backgrounds of the KKLMMT variety \cite{Kachru:2003sx}.  A recurrent
obstacle in such studies is the eta problem (familiar from
supergravity inflation \cite{Copeland:1994vg}) which essentially
pertains to the flatness of the potential.  It can be shown that
higher order corrections to the inflaton mass will give rise to an
$\mathcal{O}$(1) correction to the inflaton mass, creating a potential
that is too steep to support an inflationary phase lasting long enough
to coincide with observation.  This correction then needs to be
cancelled by terms arising, for example, from the geometric
considerations of the specific compactification manifold under
investigation \cite{Baumann:2007ah, 
 Baumann:2007np}.  The eta problem may then be circumvented, but only
in a finely tuned fashion.   

Since the eta problem is ubiquitous in inflationary models in string theory
we would like to investigate another solution, 
warm inflation (WI) \cite{Berera:1995ie}, which has been shown in
other contexts to naturally alleviate the eta problem.  
Due to
the number of fields which arise in string compactifications, a stringy 
setting provides all the requisite ingredients for warm inflation.  Although 
warm inflation may operate in any given background (including the warped 
background alluded to earlier), our specific model will be realized in a 
space with an enhanced symmetry point (ESP), leaving investigations of other
backgrounds to future works.  

An ESP is a point in field space where gauge symmetries may be
enhanced and massive fields can become massless.  This has the
interesting property that it can lead to the dynamical trapping of
moduli fields \cite{Kofman:2004yc}, giving the ESP the status of a
preferred point in field space.  The trapping mechanism has been
studied \cite{Bueno Sanchez:2006eq} as an environment where inflation
may be implemented.  Imagine a scalar inflaton  $\phi$ coupled to
another scalar field $\xi$ that becomes massless (due to its
interaction with the inflaton) at an ESP.  The particles associated
with $\xi$  will be copiously produced as the inflaton moves towards
the ESP.  The inflaton will be caught in a damped oscillation about
the ESP, giving up its to particle production.  The kinetic energy of
the oscillation will eventually fall below that of the potential,
leading to trapped inflation.  The universe may then be reheated by
the decay of a curvaton field. 

Here we propose an alternative method of inflation and reheating that
begins with moduli trapping  but concludes with 
heating through the WI paradigm
\cite{Berera:1995ie,BasteroGil:2006vr}. 
Inflationary expansion and radiation production
occur concurrently and arises dynamically in WI through the the presence
of  fields with which the inflaton interacts during inflation,
resulting in dissipation and particle production.
The presence of radiation during inflation also allows
the end of inflation to occur when the vacuum energy
density is falling faster than the radiation energy density,
so that at some point a smooth crossover occurs
from inflation to radiation domination, without
a  separate reheating phase. (For reviews of warm
inflation and its origin from quantum field theory
see \cite{Berera:2006xq}.)

We examine a generalization of the model found in \cite{Kofman:2004yc}
where we include additional scalar fields which leads to an alteration
of the trapping and provides new channels of inflaton energy
dissipation  leading to warm inflation.  Moduli fields naturally arise
in stringy contexts due to perturbations or deformations of the
background compactification geometry or the presence of branes and
fluxes.   The number of such moduli fields is determined by the
compactification data and can be quite large.  For example, in
Calabi-Yau compactifications one finds complex structure (shape) and
Kahler (size) moduli, whose multiplicities are given by the Hodge
numbers $h^{2,1}$ and $h^{1,1}$ respectively.  The presence of branes
and fluxes adds more to the field count.  It is not atypical then to
expect the number of degrees of freedom of the system to run into the
hundreds or thousands (some examples with large field numbers are
found in assisted inflation in M-theory with hundreds of  
fields \cite{Becker:2005sg}, or $\mathcal{N}$-flation which may
incorporate thousands of axion fields \cite{Dimopoulos:2005ac}).
Recently there has also appeared work \cite{Battefeld:2008py}  which
incorporates large numbers of dynamical fields in staggered inflation
models, where the inflaton fields decay after encountering a steep
drop in their potential in contrast to the warm inflation mechanism
employed here.  

Many works choose to limit modulus numbers by invoking stabilisation
mechanisms, for all but one or a few moduli at high scales.  The
resulting effective action has fewer degrees of freedom operating
during inflation, facilitating the calculation and providing less
complicated dynamics.  The naturalness of arriving at an effective
action with only a single or few  moduli can be questioned,
especially in light of the multiplicity of vacua in the string
landscape.  Here we  assume  many moduli participate during the
inflationary phase.  The moduli fields cannot be too light (the mass
should be $m \gtrsim 100$ TeV), otherwise the successful predictions
of nucleosynthesis would be jeopardized \cite{Coughlan:1983ci}. 

An investigation of trapped inflation originating in string theory has
been undertaken in \cite{Green:2009} using wrapped D4-branes in type
IIA compactifications \cite{Silverstein:2007}, as well as a scenario
involving axion moduli in a type IIB compactification on a warped
Calabi-Yau space \cite{McAllister:2008}.  Some ingredients of the warm
inflation scenario we are presenting are seen to  arise there (such as
large numbers of light fields, where, in the IIB example, the number
of light species is given by the square of the number of circuits of
the fundamental axion that can fit within the compactification, and
the number of circuits can be of $\mathcal{O}(10^3-10^4)$).  Although
we have not employed these specific examples in our present work, we
find their structure to be complementary to our approach.

We now focus on a generalization of the model of \cite{Kofman:2004yc}
\begin{eqnarray}
\label{L1}
\mathcal{L} = \frac{1}{2}\partial_{\mu}\phi\partial^{\mu}\phi + \frac{1}{2}\partial_{\mu}\chi\partial^{\mu}\chi - 2g^2\phi^2\chi^2\,,
\end{eqnarray}
where $\phi$ is the inflaton field and $\chi$ is another scalar field.  Let $\phi$ approach the origin with trajectory $\phi(t) = i\mu + vt$, where $\mu$ is the impact parameter.  At the point $<\phi> = 0$, the $\chi$ field will become massless.  It is assumed that the $\phi$ field has some time dependence which will then lead to production of $\chi$ particles that will largely take place near the symmetry point $<\phi> = 0$.  

We begin with a superpotential of the form
\begin{eqnarray}
W = \frac{1}{\sqrt{2}}(g\Phi X^2  + hXY^2)  \label{WXY}\,,
\end{eqnarray}
where the chiral superfield $X$ has $\chi$ and $\psi_{\chi}$ as its scalar and fermionic components, and the chiral superfield $Y$ has $\lambda$ and $\psi_{\lambda}$ as its scalar and fermionic components. 
Eventually one would generalize the model to many fields, since this is the generic situation provided by string theory \cite{Berera:1999wt}. 
The coupling constant $h$ is kept arbitrary.  This superpotential introduces the following terms into the action:
\begin{eqnarray}
-\mathcal{L} \supset \frac{1}{2}g^2\chi^4 +\frac{1}{2}h^2\lambda^4+2gh\phi\chi\lambda^2+2h^2\chi^2\lambda^2+2g^2\phi^2\chi^2\\\nonumber +\frac{1}{\sqrt{2}}(2g\chi\psi_{\chi}\psi_{\phi} + 2h\lambda\psi_{\lambda}\psi_{\chi}+g\phi\psi_{\chi}\psi_{\chi}+h\chi\psi_{\lambda}\psi_{\lambda})
\end{eqnarray}

We now have all of the ingredients for warm inflation to operate.
Namely, the inflaton field will excite a massive $\chi$ field which
can then annihilate either back to the inflaton
($\chi\chi\rightarrow\phi\phi$) or to the $\lambda$ fields
($\chi\chi\rightarrow\lambda\lambda$), while also directly decaying
$\chi\rightarrow\psi_{\lambda}\psi_{\lambda}$, the last two processes
will then dissipate the energy of the inflaton field.   

We will first study the effects of the new annihilation path
$\chi\chi\rightarrow\lambda\lambda$ on the trapping mechanism.  The
effect of the $\chi\chi\rightarrow\phi\phi$ reaction on the trapping
length was studied in \cite{Kofman:2004yc}.  Inclusion of the
additional $\chi\chi\rightarrow\lambda\lambda$ requires only slight
modification.  The interaction cross section for each process is given
by 
\begin{eqnarray}
\sigma_{\phi} = \frac{4g^4k'}{\pi k E^2}\,\,\,;\,\,\,\sigma_{\lambda}
= \frac{4h^4k'}{\pi k E^2}\,, 
\end{eqnarray}
where $k$ is the incoming momentum, $k'$ is the outgoing momentum, $E$
is the incoming energy, and the subscript indicates which field the
$\chi$ will annihilate to form.  The rate of loss of $\chi$ particles
is then given by the change in number density 
\begin{eqnarray}
\frac{\dot{n}(\vec{k}_1,t)}{n(\vec{k}_1,t)} = -\int
d\vec{k}_2n(\vec{k}_2,t)\frac{\sqrt{(k_1k_2)^2
    -m_{\chi}^4}}{E_1E_2}(\sigma_{\phi}+\sigma_{\lambda})\,.  \end{eqnarray} 
This can then be evaluated in the center of mass frame in the
non-relativistic limit, while ignoring the mass of the $\phi$
particle, to obtain a bound on the decay 
\begin{eqnarray}
\frac{\dot{n}(\vec{k}_1,t)}{n(\vec{k}_1,t)} \ge -\frac{8(g^4+h^4)}{\pi
  m_{\chi}^2}\int d\vec{k}_2n(\vec{k}_2,t)\,. \end{eqnarray} 
The mass of the $\chi$ field is known from the $\phi$ motion and the
total number density of $\chi$ particles produced when $\phi$ encounters the ESP, $n_{\chi}$, can be used to
find the bound 
\begin{eqnarray}
\frac{n(\vec{k}_1,t)}{n(\vec{k}_1,0)} \ge
\exp\left[-\frac{8(g^4+h^4)n_{\chi}}{\pi(\mu^2 + v^2t^2)}\right]. 
\end{eqnarray}
One then needs to calculate the energy density from the $\chi$
particles in order to find the new trapping length.  The energy is
given by 
\begin{eqnarray}
E = \int d\vec{k}n\sqrt{k^2 + 4g^2(\mu^2 + v^2t^2)}\\\nonumber \ge
n_{\chi}\sqrt{4g^2(\mu^2 +
  v^2t^2)} \exp\left[\frac{-8(g^4+h^4)n_{\chi}}{\pi\mu v}\right]\,, 
\end{eqnarray}
which leads to the new stopping length
\begin{eqnarray}
\phi_{*} = \frac{\pi^3}{\sqrt{2g^{5}}}v^{1/2} \exp\left[\frac{2\pi
    g\mu^2}{v^2}\right] \exp\left[ \frac{{4(g^4+h^4)n_{\chi}}}{\mu v}
  \right]\,. 
\end{eqnarray}
We see that the inclusion of an additional decay path for the $\chi$
particles will not ruin the trapping effect, only lengthen it with the
additional $h$-dependent term $\exp[4h^4n_{\chi}/(\mu v)]$, while it will
provide the necessary conditions for warm inflation.

In an expanding universe, once particle production stops, the amplitude
of the oscillations around the ESP will decrease like the inverse of
the scale factor \cite{Bueno Sanchez:2006eq}. Afterwards when the
kinetic energy falls below that of the potential a period of inflation
can start. With the interactions in the superpotential Eq. (\ref{WXY})
we can have warm inflation through a two-stage mechanism
\cite{Berera:2001gs}, in which the energy of the inflaton is
dissipated into radiation through the $X$ fields into the light $Y$
fields. The extra friction term in the inflaton evolution equation
induced by the dissipative dynamics translates generically into a
longer period of inflation, and therefore the possibility of having 60
$e$-folds of trapped inflation.

The dissipative coefficient has been computed 
based on the adiabatic approximation for warm inflation
quantum field theory solutions \cite{mx}.
The dominant contribution comes from the interaction of the inflaton
field $\phi$ with the scalar $\chi$ and that of $\chi$ with the scalar
$\lambda$.
The interactions
depend on the coupling ``$h$'' in the
superpotential, which previous to inflation leads to
the direct decay  path $\chi \rightarrow \psi_\lambda
\psi_\lambda$. Therefore in order to allow for trapping one has   
first to check that the $\chi$ particles do not decay before the inflaton
field has reached the turning point, such that the interaction
potential induced by them do not disappear. Due to the expansion, 
the decay takes place when the decay rate $\Gamma_\chi$ equals the
Hubble rate, so that the condition for trapping in the presence of
this direct decay channel is then: $\Gamma_{\chi_{TR}} < H_{TR}$,
where the subindex ``TR'' refers to the trapping period. It is enough 
to check this condition when the decay rate gets its larger value, at the turning
point with $\phi$ equal to the amplitude of the oscillations around the ESP. 
On the other hand, consistency of the approximations during warm
inflation requires the decay rate to be larger than the Hubble rate
during inflation, i.e.: $\Gamma_{\chi_{inf}} > H_{inf}$, with the subindex
``$inf$'' denoting now the time of inflation.  

Taking into account that we
may have many fields $X$ and $Y$, the total number of $\chi$ particles produced
during trapping is enhanced by a factor ${\cal N}_\chi$ counting the
multiplicity of the $X$ field, and the stopping length is therefore
shortened by the inverse of this factor
\be
\phi_* \simeq \frac{4 \pi^3}{g^{5/2}}v^{1/2} {\cal  N}_\chi^{-1} \,,
\label{phiTR}
\ee
The decay rate is given by
$
\Gamma_\chi = \frac{h^2}{8 \pi}{\cal N}_{dec} m_\chi \,,
$
where ${\cal N}_{dec}$, the number of $Y$ fields, 
is the number of decay channels
available in
$X's$ decay, and $m_\chi= g \phi$. During particle production,  when
the field passes the ESP and reaches the turning point, the Hubble rate is
given by the kinetic energy of the field, with $H_{TR} \simeq
v/\sqrt{6}$. Using Eq. (\ref{phiTR}) in
$\Gamma_{\chi_{TR}} < H_{TR}$, and demanding also $\phi_* < m_P$ to avoid
overshooting, the condition to avoid $\chi$ decay reads:    
\be
{\cal N}_{dec} \frac{h^{2}}{8 \pi} <  \frac{{\cal N}_\chi^2}{\sqrt{6}\pi^4} \left(\frac{g^2}{4 \pi} \right)^2
\,. \label{trappingcond}  
\ee
Taking for example ${\cal N}_\chi \gtrsim O(100)$ and ${\cal N}_{dec}
\lesssim {\cal N}_\chi$, the above condition can be easily fulfilled
with both couplings of the order of $O(1)$, without the need of imposing
any hierarchy between them. 

To check the consistency condition for warm inflation we have to take
into account that once particle production stops, the amplitude of the
oscillations decreases like the inverse of the scale factor, while the
Hubble rate goes like $H \propto a^{-2}$, and therefore before
inflation the decay rate becomes larger than the Hubble rate when: 
\be
\frac{H_{inf}}{g m_P} <   \frac{{\cal N}_\chi^2}{\sqrt{6}\pi^4} \left(\frac{g^2}{4 \pi} \right)^2 \,.
\ee
For couplings of the order of $O(1)$ this is a very mild
condition on the scale of inflation, which is naturally fulfilled. 

Therefore, by considering the presence of several moduli fields, i.e.,
large multiplicities for $X$ and $Y$ fields, trapping around the ESP
is still possible even in the presence of the direct decay rate into
fermions due to the coupling $h$, with strong couplings of
order unity. At the same time, strong
couplings and large multiplicities are required in order to have
enough dissipation for warm inflation.  
In the low $T$
regime, when $m_\chi >T$, the dissipative coefficient goes like \cite{mx}
$
\Upsilon 
\simeq C_\phi h^{4} \frac{T^3}{\phi^2} \,,
$
where $C_\phi= 0.04  {\cal N}_\chi {\cal N}_{dec}^2$. With $h^2
\simeq 4 \pi$ and multiplicities ${\cal N}_\chi\simeq 200$, ${\cal
  N}_{dec}\simeq 30$ we have $C_\phi \simeq 10^6$; by increasing the
multiplicities by a factor of say 5, we would have $C_\phi \simeq
10^8$. These are  typical values required to
realize WI within the low $T$ approximation
\cite{BasteroGil:2006vr,hilltop}, which can be accomplished   
quite naturally in $SO(10)$ or $E_6$ grand unification models and in
models where the inflaton interacts with many massive 
string modes causing dissipation of vacuum energy i.e., warm
inflation, like in \cite{Berera:1999wt}. Moreover  a coupling $g^2/(4
\pi)\gtrsim 0.3$ ensures that the bound on Eq. (\ref{trappingcond}) is
fulfilled. 

As a specific example, let us assume that the ESP generates a maximum
at vanishing field value in the scalar 
potential, and inflation proceeds after trapping with the 
field rolling away from the origin 
i.e., hilltop inflation. 
The simplest inflationary potential to  parametrize this scenario would be
$
V = V_0 - \frac{1}{2} |m_\phi|^2 \phi^2 \,. 
$
During hilltop warm inflation, the ratio of the dissipative
coefficient to the Hubble rate $Q=\Upsilon/(3H)$ decreases, but $T$ and
$m_\chi/T$ increase, so once in the low $T$ regime the approximation
holds until the end of inflation and it can be studied analytically
\cite{hilltop}. 
For example, with
a value $C_\phi \simeq 10^6$ we
can have 40-60 e-folds of inflation in the weak dissipative regime
with $Q< 1$, and a subplanckian value of the field up to the end when 
$V_0^{1/4} \lesssim 10^{-3}m_P$. The prediction for the spectral index
$n_S$ at lowest order in the slow-roll parameters is the same than in
hilltop cold inflation, $n_S \simeq 1 + 2 \eta_\phi$, so that it will be
within the observed range for a mass parameter such that $|\eta_\phi
|\simeq |m_\phi^2|/(3H^2) 
\sim 2\times 10^{-2}$. On the other hand, by increasing the
multiplicities such that $C_\phi \simeq 10^8$, we can have the 40-60
e-folds of inflation in the strong dissipative regime with $Q > 1$,
and get the right predictions for the primordial spectrum for larger
values of the mass and $|\eta_\phi|$. With $|\eta_\phi|
\simeq 1$ we have $n_S \simeq 0.94$, and  $V_0^{1/4} \lesssim 10^{-4}m_P$ ensures
that the field at the end is still below the Planck scale. Restricting the
height of the potential to retain $\phi < m_P$ during inflation, we
implicitly restrict the amount of radiation dissipated during warm
inflation (although the radiation energy density increases), such that it is
still subdominant at the end of the inflation. By further increasing
the dissipative parameter so that $C_\phi \gtrsim 10^9$, and
at the same time, the inflaton mass up to values $|\eta_\phi| \gtrsim O(100)$,
we recover the possibility of a smooth transition from inflation to
radiation domination at the end of warm inflation with $V_0 \simeq
10^{-4} m_P$, and $n_S \simeq 0.95$, i.e., well within the
observational range. In this case multiplicities and couplings of the order of  
${\cal N}_\chi\simeq 10^3$, ${\cal N}_{dec}\simeq 350$, $h^2
\simeq 4 \pi$ are required, but still with $g^2/(4 \pi) \gtrsim 0.2$
the bound in Eq. (\ref{trappingcond}) for trapping is fulfilled.

Inflation is a successful scenario in need of a theoretical home, with
string theory being the leading candidate.  This decade has seen an
explosion of ideas on just how this may be accomplished.  One obstacle
which is ubiquitous in the string theory approach is the eta problem.
There currently is much work going towards its solution but some
residual fine tuning appears to remain.  We have examined an
alternative approach towards rectifying the eta problem, namely
invoking the warm inflation scenario, which changes the nature of, and
subsequently solves, the inherent eta problem.  String theory seems to
be a natural place for the warm inflationary mechanism to operate due
to generic presence of multitudes of moduli fields.  Here we have
demonstrated a possible example of warm inflation in the context of
trapped inflation, a scenario which is interesting in its own right
from a field theoretic viewpoint, but has also been demonstrated to
arise in string contexts.  Although we are looking at only this corner
of the vast landscape of string theory, we believe this to be an
instructive example which may be implemented in other inflationary
models in string theory.  Since warm inflation can provide a natural
solution to the eta problem, which has been a major barrier to string
model building, and since strings possess a large number of fields,
which is an important ingredient for most warm inflation models, we
believe warm inflation may have a natural setting in string theory.
We believe that this is a step towards the warm inflation scenario,
and its attendant attractive properties, finding a home in the string
realm.  

\vspace{-.07in}

MBG is partially supported by MICINN grant FIS2007-63364 and JA
grant FQM101. AB is partially supported by STFC. 
JBD and TWK were supported by US DoE grant
DE-FG05-85ER40226. TWK  thanks the Aspen Center for Physics for
hospitality while this research was in progress.

\end{document}